\documentstyle[preprint,aps,prd]{revtex}

\input epsf

\preprint{Sub: 25 Oct 95; Acc: 3 Jan 96}

\begin{document}

\baselineskip14pt

\def\spose#1{\hbox to 0pt{#1\hss}}
\def\simlt{\mathrel{\spose{\lower 3pt\hbox{$\mathchar"218$}}
     \raise 2.0pt\hbox{$\mathchar"13C$}}}
\def\simgt{\mathrel{\spose{\lower 3pt\hbox{$\mathchar"218$}}
     \raise 2.0pt\hbox{$\mathchar"13E$}}}
\def\degrees{\hbox{${}^\circ$\hskip-3pt .}}
\def\hMpc{h^{-1}{\rm Mpc}}

\title{Cosmic Confusion and Structure Formation}
\author{Martin White}
\address{Enrico Fermi Institute, University of Chicago\\
5640 S.~Ellis Ave, Chicago~IL 60637}

\maketitle

\begin{abstract}
We present a re-analysis of the cosmic confusion hypothesis, elucidating the
degree to which ``confusion'' can be expected to hold in a class of flat,
adiabatic models.  This allows us to devise a simple and accurate fitting
function for the height of the first peak in the radiation power spectrum in
a wide range of inflationary inspired models.
The range preferred by current data is given, together with a discussion of
the impact of measurements of the peak height on constraining models of
structure formation.
\end{abstract}


 
\section{Introduction}

The picture of structure formation through gravitational instability now
appears to be well established.
Initial fluctuations grow through gravitational instability to form the
structures which we observe today, and leave their imprint at redshift
$z\sim1000$ in the cosmic microwave background (CMB) radiation anisotropy.
Within this framework any theory which purports to explain the large-scale
structures we observe today, must simultaneously fit the increasing amount
of data on CMB anisotropies.
The combination of these two constraints is especially powerful, since they
probe a large lever arm both in scale ($k\simlt10^{-4}\hMpc$ to
$k\simgt1\hMpc$) and time ($z\simgt1000$ to $z\simeq0$).

\begin{figure}
\begin{center}
\leavevmode
\epsfysize=12cm \epsfbox{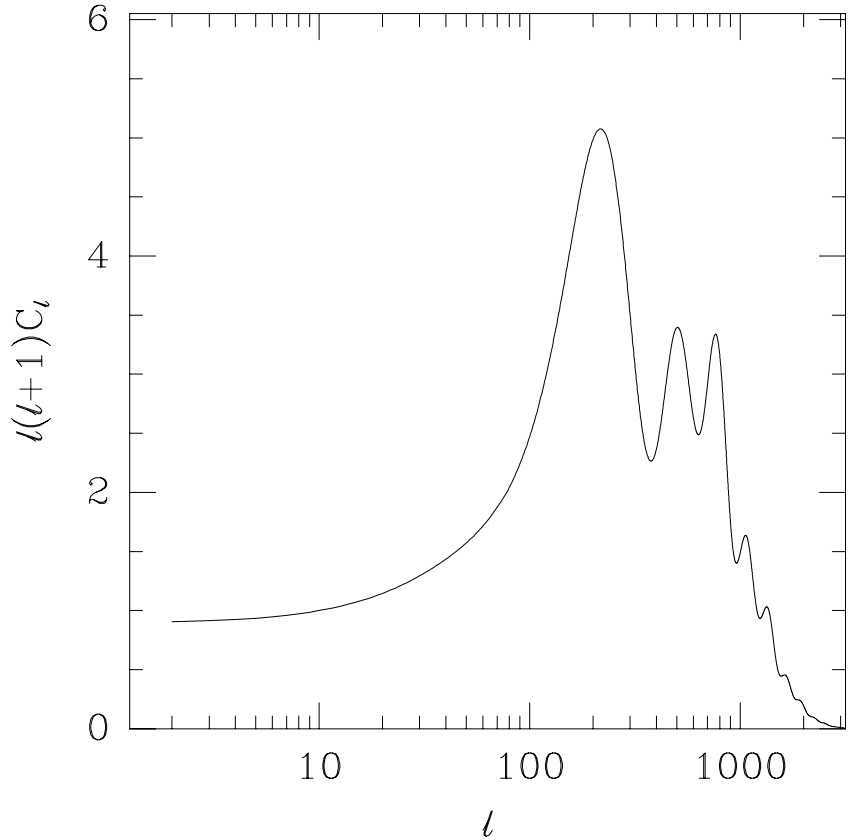}
\end{center}
\caption{The CMB anisotropy power spectrum, $\ell(\ell+1)C_\ell$ vs $\log\ell$,
for the ``standard'' Cold Dark Matter model with $\Omega_0=1$, $\Omega_B=0.05$,
$n=1$ and $C_2^{(T)}=0$.  The vertical axis is approximately the power per
logarithmic interval in multipole $\ell$, which probes angular scales
$\theta\simeq\ell^{-1}$.
For experiments on angular scales $\theta\simgt0\degrees5$ we are interested
only in the rise into the first peak (at $\ell\simeq220$).
\label{fig:cl_ref}}
\end{figure}

Calculations of both CMB anisotropy spectra and matter power spectra are
now well developed for inflationary models, with the former approaching 1\%
precision \cite{Sug,HSSW}.  (In Fig.~\ref{fig:cl_ref} we show the anisotropy
power spectrum for the ``standard'' CDM model, see \cite{Sel,echoes,SSW,HuSug}
for a discussion of the physics behind these anisotropies).
This kind of precision in theoretical predictions is necessary to interpret
the results envisioned from a future satellite CMB mission, which could
accurately measure the power spectrum over a broad range of scales, allowing
one to measure most of the standard cosmological parameters with unprecedented
accuracy \cite{echoes,JunKKS}.
However such a measurement is still (optimistically) several years away, and
in the meantime the data on CMB anisotropies continues to be amassed at a
steady rate.  For the interpretation of current data, such a high level of
precision is not necessary, and semi-analytic techniques exist which can give
spectra accurate to $\sim10\%$ \cite{HuSug}.

An important constraint on structure formation models can be obtained without
even this amount of effort however.
It has been pointed out \cite{Bondetal,HuSug,Sel} that for measurements
of the spectrum constrained to scales of $\sim0\degrees5$ or above, the
anisotropy power spectra exhibit a degeneracy in parameter space, which can
be both a hindrance and a boon.
This degeneracy has been dubbed ``cosmic confusion'' \cite{Bondetal} as it
provides a limit to how well cosmological parameters can be extracted from
measures of the CMB power spectra on scales larger than $0\degrees5$.

\begin{figure}
\begin{center}
\leavevmode
\epsfysize=12cm \epsfbox{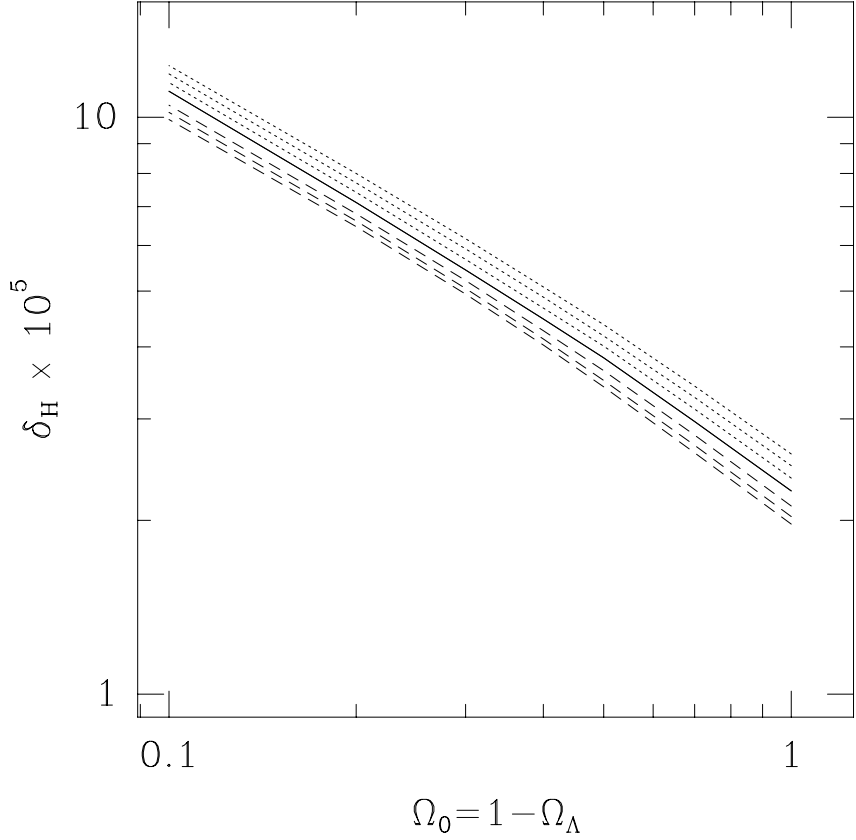}
\end{center}
\caption{The dimensionless amplitude of the matter power spectrum at horizon
crossing, $\delta_H$, vs.~matter density $\Omega_0=1-\Omega_\Lambda$ for
$n=0.85,0.9,0.95,1.00$ as fit to the 2-year {\sl COBE} data.
$\delta_H$ is defined in terms of the r.m.s.~fluctuation per logarithm in
wavenumber, $k$, by $\Delta^2(k)=k^3P(k)/(2\pi^2)=\delta_H^2(k/H)^{3+n}T^2(k)$.
The solid line is a scale-invariant spectrum: $n=1$.
The dotted lines show values for lower $n$ (lowest $n$ is highest line) with
no tensor contribution.
The dashed lines show the same range of $n$ (now lowest $n$ is lowest line) for
models with a tensor contribution as predicted by power-law inflation.
These results include the dependence of the tensor-scalar ratio on
$\Omega_\Lambda$ that was incorrectly omitted in Ref.~18.
\label{fig:deltah}}
\end{figure}

Notice the promise this degeneracy has in simplifying the interpretation
of degree scale CMB data.
As noted by Dodelson \& Kosowsky \cite{DodKos}, if this degeneracy holds
it is possible to characterize the constraints on a wide range of models in
terms of two parameters: an overall amplitude and the value of the degenerate
combination of parameters $\nu$ (see \S\ref{sec:confusion} for details.  A part
of this approximate degeneracy was implicitly used in \cite{SSW} to constrain
the spectral slope and ionization history from the peak height).
The amplitude of the {\it potential} fluctuations, measured on the largest
scales by {\sl COBE}, has become the preferred method for normalizing theories
of large-scale structure \cite{cobenorm}.
(We show an example of this in Fig.~\ref{fig:deltah}, where the
dimensionless amplitude of the matter power spectrum at horizon crossing,
$\delta_H$, is shown vs.~$\Omega_0=1-\Omega_\Lambda$ and spectral tilt $n$.)
The value of $\nu$ will constrain other parameters on which the predictions of
the theory depend.  Furthermore, to the extent that cosmic confusion holds,
generating spectra for a wide range of parameters can be reduced to simple
modifications of a standard spectrum (however with the advent of semi-analytic
techniques for quickly generating spectra over the entire angular range, this
may be of limited utility).
An obvious application of this is a fitting formula for the height of the
peak in the power spectra at $\ell\simeq220$.  The current data provide limits
on the height of the peak (see e.g.~\cite{SSW}), which are becoming stronger.
A quick method of estimating the height of the peak in the CMB spectrum
can allow one to estimate the constraints from degree-scale measurements
quickly and efficiently over a broad range of parameter space.

The outline of this paper is as follows.  In \S\ref{sec:confusion} we
discuss how well cosmic confusion works, and slightly modify the original
formulation \cite{Bondetal}.  \S\ref{sec:reion} discusses the special case
of late reionization in processing the spectra.  In \S\ref{sec:fit} we use
a compilation of CMB data to provide a limit on the peak height and discuss
implications for models of structure formation.
\S\ref{sec:conclusions} presents the conclusions.

\section{Cosmic Confusion} \label{sec:confusion}

In this section we would like to make some comments about, and minor
improvements upon, the degeneracy of CMB spectra with respect to certain
parameters which has been dubbed ``cosmic confusion'' \cite{Bondetal}.
Specifically, it is claimed that for a range of parameters the CMB anisotropy
spectra are degenerate in
\begin{equation}
\widetilde{n} = n   - 0.28\ln(1+0.8r)
                    - 0.52\left( \Omega_0^{1/2}h - 0.5 \right)
                    - 0.00036 z_R^{3/2}
\end{equation}
where $n$ is the (scalar) spectral index, $r$ is the ratio of tensors to
scalars (see below), $h$ is the Hubble constant in units of
$100{\rm km}\,{\rm s}^{-1}\,{\rm Mpc}^{-1}$ and $z_R$ is the reionization
redshift (see \S\ref{sec:reion}).
The degeneracy is supposed to hold up to the first peak in the power
spectrum, $\ell\simeq 220$, though some spectra can be degenerate over an
even larger range of $\ell$ \cite{Bondetal,Crossroads}.

To start let us take as our guiding principle the following conjecture:

\begin{itemize}
\item{} For reasonable input parameters all CMB anisotropy spectra with
the same peak-to-plateau ratio will be approximately degenerate.
\end{itemize}

The interpretation of this statement clearly depends upon what range of
cosmological models are adopted, over what range of scales the degeneracy
is required to hold (not beyond the first peak), and the precise meaning
of ``approximately''.  We will discuss all of these issues below.

Inflation in its most generic form predicts a flat universe, so in this
paper we will restrict ourselves to models with vanishing spatial curvature.
Several questions relating to open inflation are still unresolved,
for example the predictions for super-curvature modes, spectral tilt and
tensor modes\footnote{The contribution to the large-angle CMB anisotropy from
tensor perturations will probably be small in open models, since a
combination of the {\sl COBE} and cluster abundance normalizations predicts
the spectral index $n$ to be very close to (or greater than) 1.}.
While for any given model the calculations of the power spectrum are
straightforward \cite{open,SSW}, we feel that these uncertainties make
fitting to spectra premature for these models.

The simplest implementation of cosmic confusion is in terms of an apparent
spectral index, which we shall call $\nu$ (this is related to the
$\widetilde{n}$ of \cite{Bondetal} by $\nu\simeq\widetilde{n}-1$).
As our base model we shall adopt `standard' CDM ($h=0.5$, $\Omega_0=1$,
$\Omega_\Lambda=0$, $n=1$, $C_2^{(T)}=0$) for which we define $\nu=0$.
This spectrum is shown in Fig.~\ref{fig:cl_ref} and tabulated in
Table~\ref{tab:scdm}.
The spectrum for a model whose parameters combine to give $\nu$ is then
related to that of standard CDM by the simple formula
\begin{equation}
D_\ell(\nu) = \left({\ell\over 10}\right)^\nu\ D_\ell(\nu=0)
\label{eqn:tiltapp}
\end{equation}
where $D_\ell\equiv\ell(\ell+1)C_\ell$ for notational convenience.
We have chosen to pivot around $\ell=10$ as this is approximately the
pivot point of the {\sl COBE} data \cite{Goretal}.
Alternatively one could use $\sigma(10^{\circ})$ \cite{Bondetal}, which
turns out to be almost the same for the range of spectra under consideration.

To understand the expression for $\nu$ in terms of the parameters of
the model, we need only consider the effect of changing each parameter
on the peak height of the model.
We will now consider each of these effects in turn, concentrating first
on those parameters for which the conjecture works well and then turning
to those for which the deviations are larger.

As discussed in \cite{HSSW} changing the spectral index, $n$, of the
primordial spectrum ($\propto k^n$) is equivalent to multiplying the
anisotropy spectrum by $\ell^{n-1}$ as in Eq.~\ref{eqn:tiltapp}.  For
$n\ge0.8$ the worst deviations induced by this approximation occur around
$\ell=30$--50, and are $\simlt10\%$.
Hence tilting the model away from scale invariance changes $\nu$ by
$n-1$.

Associated with tilt is the possibility that some of the CMB anisotropy
comes from long-wavelength, inflation-produced gravitational waves (tensor
fluctuations).
The fraction, $r$, of tensors is usually quoted in terms of the contribution
to the quadrupole: $r\equiv C_2^{(T)}/C_2^{(S)}$.
Unfortunately due to the decaying potentials in a $\Lambda$ model the scalar
quadrupole is very sensitive to $\Omega_\Lambda$ \cite{SacWol,KofSta} for a
fixed ``initial'' power spectrum.  The tensor quadrupole is much less sensitive
\cite{TWL} and hence the ratio has a strong (artificial) $\Lambda$
dependence\footnote{This dependence was originally neglected in \cite{WhiBun}
thus overestimating the damping effect of tensors for the tilted models.}
\cite{Knox}.
We choose instead to write $r=1.4 C_{10}^{(T)}/C_{10}^{(S)}$, where the
prefactor has been chosen to make the definitions agree for $\Omega_0=n=1$.
The $\ell=10$ mode is not as strongly affected by the decaying potentials
\cite{HuWhi} making this a more robust measure of the underlying tensor/scalar
ratio.
Since the tensor spectrum damps rapidly on scales smaller than the horizon at
last scattering \cite{TWL,CBDES} the effect of introducing a tensor component
is to increase the plateau by $(1+0.76r)$ while leaving the peak unchanged.
[The factor 0.76 is roughly $D_{10}^{(T)}/D_2^{(T)}$ for a scale-invariant
spectrum with $\Omega_0=1$.]
An equivalent reduction in the peak height relative to the plateau can be
accomplished by reducing the spectral index
(from $\ell=10$ to $\ell\simeq220$) by,
\begin{equation}
\nu \ni -\ln(1+0.76r) / \ln(220/10) \simeq -0.32\ln(1+0.76r)
\end{equation}

\begin{figure}
\begin{center}
\leavevmode
\epsfysize=12cm \epsfbox{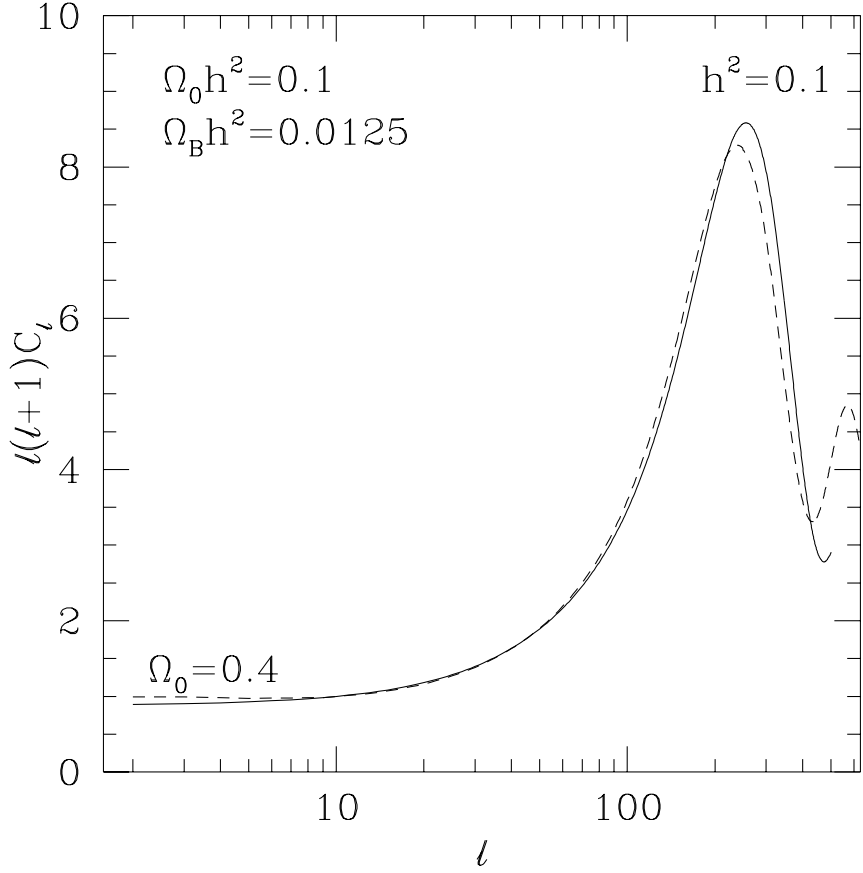}
\end{center}
\caption{The CMB anisotropy power spectra for two models with the same
$\Omega_0h^2=0.1$ and $\Omega_Bh^2=0.0125$.  The solid line is for
$\Omega_0=1$ and $h^2=0.1$, while the dashed line is
$\Omega_0=1-\Omega_\Lambda=0.4$ and $h^2=0.25$.  Notice that the spectrum
is not degenerate in $\Omega_0h^2$ and $\Omega_Bh^2$ at the 5--10\% level,
even up to the first peak.
\label{fig:cl_compare}}
\end{figure}

In the semi-analytic, two-fluid models of CMB anisotropies
\cite{DZS,JKNN,WSS,Sel,HuSug}, the height of the first peak depends primarily
on the combinations $\Omega_Bh^2$ (which sets the speed of sound in the coupled
baryon-photon fluid) and the redshift of matter-radiation equality.
The latter scales with $\Omega_0h^2f_{\gamma}$, where
$f_\gamma\equiv1.68\,\rho_\gamma/\rho_{\rm rad}=3.36/g_{*}$ is the fraction
of the radiation made up of photons, scaled to be 1 in the standard model
with 3 massless neutrino species.
In models in which more relativistic species are included, such as decaying
neutrino models \cite{decneut}, $f_\gamma$ can be reduced
leading to a larger peak height, just as if $\Omega_0$ were lowered.
As a warning we point out that this `degeneracy' is not exact, due to the
effect of $\Omega_\Lambda$ at low redshift (which is included implicitly in
the formulation of the semi-analytic models \cite{HuSug}).
In Fig.~\ref{fig:cl_compare} we show spectra for two models with the
same (reasonable) values of $\Omega_0h^2f_\gamma$ and $\Omega_Bh^2$,
which differ even up to the first peak by 5--10\%.
Specifically the models have $\Omega_0h^2=0.1$ and $\Omega_Bh^2=0.0125$,
the first with ($\Omega_0$,$h^2$)=(0.4,0.25) and the second with
(1,0.1).  As one can see from the figure, the exact height and position of
the first peak are complicated functions of the cosmological parameters!
The reason for the breaking of the degeneracy in this particular case is
that the large-angle temperature fluctuations (from decaying potentials at
$\Lambda$ domination) and the projection of physical scales to $\ell$ both
depend on $\Omega_\Lambda$, not just $(1-\Omega_\Lambda)h^2$.

The original statement of cosmic confusion was meant to hold for models
with $\Omega_Bh^2=0.0125$, meaning that models with equal
$\Omega_0h^2f_\gamma$ would be degenerate, and furthermore that they would
coincide with models with some $n\ne1$.
For models with scale invariant initial perturbations, $\Omega_0=1$,
$\Omega_Bh^2=0.0125$ and $h=0.3$ to 0.75, we find that the height of the
peak (relative to $D_{10}$) is very well fit by
\begin{equation}
\left( {D_{\rm peak}\over D_{10}} \right) \propto h^{-1.19} \quad .
\end{equation}
We note that $\ell_{\rm peak}$ ranges from 260 to 200 over this range of $h$,
since a changing sound speed causes the sound horizon at last scattering
to subtend a varying angular scale \cite{Sel,HuSug}.
This small movement of the peak should not matter if we redefine
``cosmic confusion'' to hold when averaged over a reasonably broad window
in $\ell$, as is the case in most experiments to date.
Under the ``confusion'' assumption then we predict that
\begin{equation}
\nu \ni -0.37 \ln(2h)
\end{equation}
which is roughly equivalent to the original statement given in
\cite{Bondetal}.
Both this approximation, and the original statement \cite{Bondetal}, give
$\simlt10\%$ deviations in $D_\ell$ for $10\le\ell\le250$ and
$0.35\le h\le0.75$.
[The percentage deviations for $\ell<10$ can be larger than 10\%, but
here cosmic variance is also larger.]

Similarly we find that the variation with $\Omega_0=1-\Omega_\Lambda$ at
fixed $\Omega_Bh^2$ and $h$ works at the $\sim10\%$ level in power
for $\ell\ge10$ and $\Omega_0\ge0.4$, with
\begin{equation}
\nu \ni -0.16 \ln(\Omega_0) - 0.15\ln(f_\gamma)
\end{equation}
This is not exactly what we would have expected based on a degeneracy
in $\sqrt{\Omega_0f_\gamma}h$: the scaling with $\Omega_0$ is slightly
weaker, and that with $f_\gamma$ weaker still.
[This can be traced to the same non-degeneracy mentioned above and illustrated
in Fig.~\ref{fig:cl_compare}.]

Note that at present $\Omega_Bh^2$ as determined by Big Bang Nucleosynthesis
(BBN) is uncertain to a factor of $\sim2$ (c.f.~the Hubble constant!)
\cite{CopST}.
Thus we are compelled to study the variation of the peak height with the
sound speed, or $\Omega_Bh^2$, which was not included in the original
statement of cosmic confusion \cite{Bondetal}.
For $\Omega_0=1$ and $0.01\le\Omega_B\le0.10$ the height of the peak is
relatively well fit by an exponential in $\Omega_B$ for $h=0.5$ and less
well fit by an exponential for $h=0.8$.
Unfortunately the slopes of the fits for these two cases (which have
different $\Omega_0h^2$) disagree:
\begin{eqnarray}
\ln\left( {D_{\rm peak}\over D_{10}} \right)
&\simeq 19\Omega_Bh^2 + 1.388 & \quad\mbox{for}\ h=0.5\\
&\simeq 23\Omega_Bh^2 + 0.824 & \quad\mbox{for}\ h=0.8
\end{eqnarray}
As a compromise then we take the average of the two coefficients to arrive
at our dependence of $\nu$ on $\Omega_Bh^2$
\begin{equation}
\nu \ni 6.8 (\Omega_Bh^2 - 0.0125)
\end{equation}
which holds relatively well over the range preferred by BBN:
$0.01\le\Omega_Bh^2\le0.02$ \cite{CopST}.

In summary then we can define a ``spectral tilt'' $\nu$ through
\begin{equation}
\begin{array}{lcc}
\nu &\equiv& n-1 - 0.32\ln(1+0.76r) + 6.8( \Omega_Bh^2-0.0125 ) \\
 &&     -0.37\ln(2h) - 0.16\ln(\Omega_0) - 0.15\ln(f_\gamma)
\end{array}
\label{eqn:nudef}
\end{equation}
For a large range of $\Omega_{\rm tot}=1$ CDM models, we find with this $\nu$
that ``cosmic confusion'' holds at the $\sim10\%$ level in power (5\% in
temperature) up to $\ell\sim200$ for those models with
$\Omega_0\ge0.4$, $h\ge0.4$ and $0.01\le\Omega_Bh^2\le0.02$
(we discuss $\tau>0$ in \S\ref{sec:reion}).  This deviation usually
occurs at $\ell\sim30$--50 and some of it can be attributed to the
approximation for tilt of the primordial spectrum that we are using
(Eq.~\ref{eqn:tiltapp}).
For low-$\Omega_0$ or $h$ models the shape of the rise into the peak and the
position of the peak are sufficiently different that a tilted model is not a
good fit (though our fit is a good approximation to the height of the peak
even for $\Omega_0\sim0.3$).
On the other hand for some models in the range mentioned the degeneracy is
good to 1\% in power for $10\le\ell\le200$.
The logarithmic dependence of $\nu$ on $\Omega_0h^2f_\gamma$ differs from
the linear dependence of $\widetilde{n}$ on the same quantity found by
\cite{Bondetal}.  Both approximations give similar ``worst fit'' $D_\ell$, but
the logarithm is a better fit to the height of the peak ($D_{220}$ relative to
$D_{10}$) over the range of models considered; it works to better than 5\%.

With the rapid progress being made in measurements of the CMB anisotropy,
it is likely that we will need to do better than 10\% in the very near future.
A more complicated fit or a better treatment of spectral tilt could alleviate
matters slightly.  However, bearing in mind that the position of the peak also
changes with the input parameters and that the degeneracies
(e.g.~$\Omega_0^{1/2}h$) built into ``confusion'' themselves only hold at the
5--10\% level, one is lead to conclude that ``one number'' summaries of the
CMB data are becoming a thing of the past.
Of course, Eq.~\ref{eqn:nudef} is still useful as a quick-and-dirty method
of estimating the height of the peak (to 5\%) in the power spectrum for quite
a wide range of parameters.
This can be useful for narrowing the large parameter space down to a smaller
region which can be searched more carefully (as one often uses linear theory
estimates in large-scale structure work).
If the experiments are chosen to probe scales for which ``confusion'' works
well, then constraints on $\nu$ will still encode much of the information that
the CMB has for large-scale structure.

As more experiments probe scales ``beyond the first peak'' ($\ell>200$),
the utility of cosmic confusion as a summary of {\it all} CMB constraints will
decrease.  In these cases one must search a multi-dimensional parameter
space and compute the full CMB spectrum either by numerical evolution of
the coupled Boltzmann equations or fast semi-analytic methods which can be
accurate to $\simlt10\%$ (e.g.~\cite{HuSug}).  
Moving beyond this degeneracy will allow us to obtain more information about
the cosmological parameters from study of the detailed structure of the
power spectrum.

\section{Reionization} \label{sec:reion}

In the original statement of cosmic confusion \cite{Bondetal}, there was
a term for the reionization redshift.
Reionization in the adiabatic models of interest is likely to occur very
rapidly \cite{reion} and at relatively low redshift, $z_R$.  Since $z_R$
the universe has probably been fully ionized.  We show in Table~\ref{tab:zr}
a ``reasonable'' estimate of the reionization redshift for some models by
way of example.  These determinations rely on several assumptions and so
should only be taken as illustrative. The results are also very sensitive to
the value of $\Omega_Bh^2$ and $\sigma_8$ chosen.

In such a scenario, the amplitude of the fluctuations on small scales
is reduced by $e^{-2\tau}$ \cite{WSS,HuSS,HuWhi}, where
\begin{equation}
\tau=0.035\,{\Omega_B\over\Omega_0}h\,x_e
     \left[ \sqrt{\Omega_0(1+z_R)^3+1-\Omega_0}-1 \right]
\end{equation}
(for $\Omega_0+\Omega_\Lambda=1$) is the optical depth to Thomson scattering
from $z=0$ to $z_R$ (also shown in Table~\ref{tab:zr}).
Note that $\tau$ depends not only on $z_R$ but also on $\Omega_B$ and $h$,
which was neglected in the original treatment of cosmic confusion
\cite{Bondetal}.
(For the accuracy to which the spectrum depends on and scales with $\tau$ see
\cite{HSSW}.)
In addition new fluctuations are generated on larger angular scales for
$\tau\simgt0.1$ \cite{WSS,SugVitSil}.
Reionized spectra are {\it not} well described by a simple tilting of the
sCDM spectrum.  An approximation which has the correct asymptotic forms
\cite{HSSW} is to multiply the spectrum by $\exp[-2\tau(z_\ell)]$, where
$\tau(z')$ is the optical depth from $z=0$ to redshift $z=z'$ and $z_\ell$
defines a mapping from angles to redshift (e.g.~the redshift at which the
horizon subtends an angle $\ell^{-1}$).
While this approximation has the right asymptotic form, it fails to be
a good approximation to a reionized spectrum for $\tau\simgt0.1$ since it
does not take into account the fluctuations generated on the new last
scattering surface.  These fluctuations typically do not extend to
$\ell>100$, so if one is interested only in scales near the peak of
the spectrum a reduction of $\exp[-2\tau]$ is appropriate.

Because reionization is such an important part of interpreting CMB anisotropy
measurements in the context of cosmological models, we will defer a more
detailed discussion of reionization to a future paper \cite{WhiHu}.

\section{Current limits on $\nu$} \label{sec:fit}

In this section we discuss what we can infer about $\nu$ from current
observations of CMB anisotropy.  We shall use the data tabulated in
\cite{SSW} since the newer data is (mostly) not yet available.  As discussed
in \cite{SSW,WSSD} and below, the constraints even without including the
new data are very interesting for large-scale structure modelling.

\begin{figure}
\begin{center}
\leavevmode
\epsfysize=12cm \epsfbox{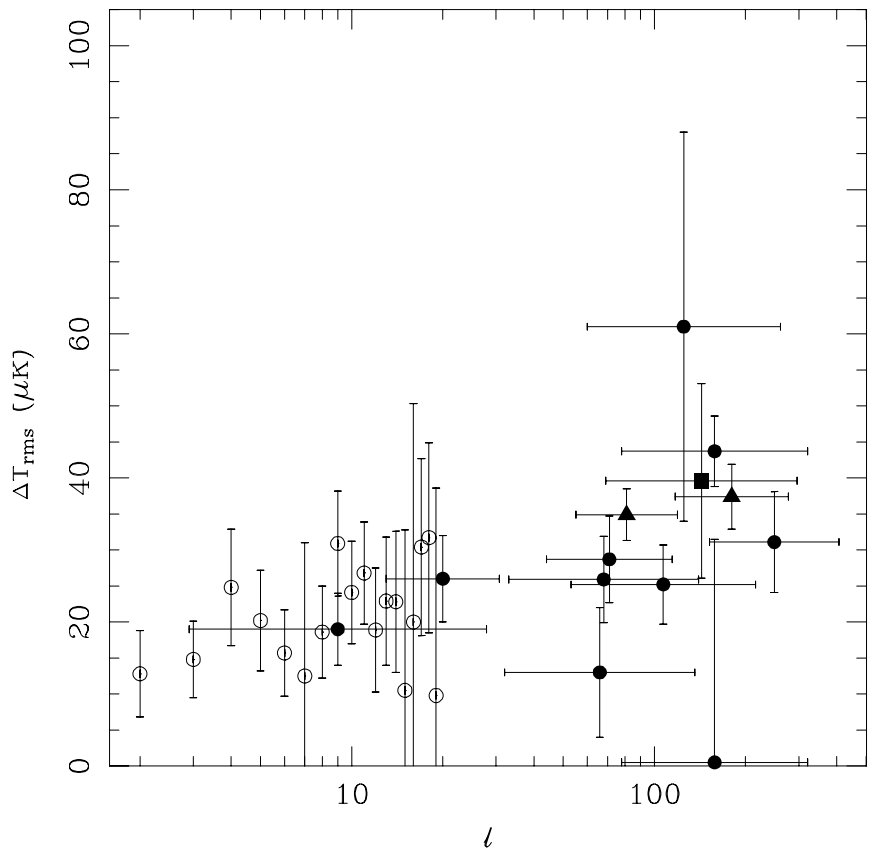}
\end{center}
\caption{The current situation with respect to CMB data, which provide
constraints on $\nu$.  The open circles come from a maximum likelihood fit to
the {\sl COBE} 2-year data (Ref.~35) and are highly correlated.  The filled
circles are the experiments tabulated in Ref.~5.  The errors are $\pm1\sigma$,
but the upper limit is 95\%CL.  The horizontal bars on each point show the
range of scales probed by each experiment.  (See Ref.~5 for more details.)
The points which are labelled by squares and triangles are not included in the
analysis: see text for discussion.
\label{fig:the_data}}
\end{figure}

Several issues regarding foreground contamination and possible systematic
errors intrude in the analysis of current CMB data.
It was shown in \cite{DodSte} that for the experiments dominating the fit
near the peak, removing the foregrounds does not increase the error bars by
more than $\sim10\%$.  To account for foreground removal we have
(conservatively) multiplied the errors in \cite{SSW} by 10\%
(except for {\sl COBE}).
Note that since \cite{SSW}, several experiments have reproduced earlier
observations, indicating that systematic errors are not as severe as might
have been thought.  In one case however, that of the MSAM experiment
\cite{MSAM}, one of the two channels (the ``single-difference'') showed
a discrepancy.  To be very conservative we have dropped this point from
our analysis (shown in Fig.~\ref{fig:the_data} as the solid square at
$\ell\simeq150$) and updated the ``double-difference'' point to include
the new data (which is slightly lower than the older data).
Also we do not include the new data from the Python experiment (though we
show it in Fig.~\ref{fig:the_data} as the solid triangles with the small
error bars \cite{Python}) since the points are correlated in an unknown way.

As it is not the purpose of this paper to revisit the data analysis,
especially without access to the latest data, we shall use the data set as
tabulated in \cite{SSW}, with the simple modifications discussed above.
One point deserves special mention.  In \cite{SSW} we used symmetric error
bars on all of the points.  This is a conservative method if the inference
is the presence of a peak, since most of the observations have skew positive
error bars.  However this leads to a slightly stronger than warranted
{\it upper} limit on the height of the peak, and a slightly weaker lower
limit.  As most of the skewed likelihood functions have not been published
in tabular or graphical form we have not tried to correct for this bias.
Since for most of the models under consideration (especially the
models with high $\Omega_\Lambda$) the large-scale structure data is best
fit with a tilt ($n<1$), this treatment remains the most conservative.

\begin{figure}
\begin{center}
\leavevmode
\epsfysize=12cm \epsfbox{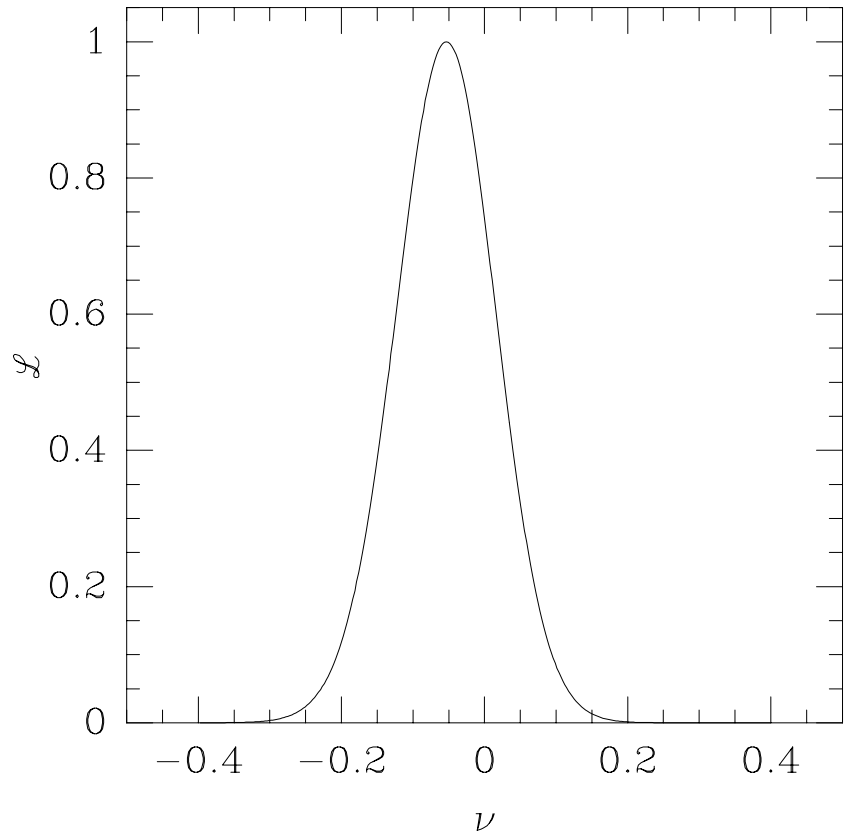}
\end{center}
\caption{The likelihood function for $\nu$, adapted from Ref.~5.  This
likelihood has been integrated (or marginalized) over the normalization
$D_{10}^{1/2}$, which is well fixed by {\sl COBE}.
\label{fig:nulik}}
\end{figure}

A fit to the data as in \cite{SSW}, but with spectra generated from
Eq.~\ref{eqn:tiltapp}, gives a likelihood function for $\nu$ shown in
Fig.~\ref{fig:nulik}.  Here we have integrated, or marginalized, over
the normalization $D_{10}^{1/2}$, which is well fixed by {\sl COBE}
\cite{cobenorm}.  The mean and standard deviation are $\nu=-0.05\pm0.07$.
The absolute goodness of fit to both the {\sl COBE} data alone \cite{WhiBun}
and the other data \cite{SSW} show no indications that the fit should be
suspect for purely {\it statistical} reasons.

Notice that this result has several immediate implications for structure
formation models.  That the preferred peak height is near that of a CDM
model with scale invariant initial conditions puts a {\it lower} limit on
the amount of tilt which can be accommodated \cite{SSW,WSSD}.
The lower limit clearly depends on the values of $\Omega_0$, $h$ and
$\Omega_Bh^2$ assumed, with lower $\Omega_0$ and $h$ acting to loosen the
constraint and lower $\Omega_Bh^2$ acting to strengthen it.
The probability of early reionization \cite{reion} in these models
will only tighten the lower limit on $n$.

The increased height of the peak in low-$\Omega_0$ models or models with
$f_\gamma<1$ can cause conflict with the {\it upper} limit on $\nu$, unless
the models are tilted, the baryon fraction is lowered or there is some
reionization.  We note in passing that for $\Lambda$CDM, some tilt is
probably necessary to provide a good fit with the large scale structure data
in any case \cite{SSW,WhiBun,KlyPriHol,WhiSco}.
In general, while the degree scale data provide a strong constraint on models
with high $\Omega_0$, the possibility of early reionization (the degree scale
power is exponentially sensitive to the very uncertain $z_R$) and the large
uncertainty in $\Omega_Bh^2$ from BBN significantly weaken the constraints
from this data for low $\Omega_0$ or $f_\gamma$ models.

\section{Conclusions} \label{sec:conclusions}

In this paper we have compared the accuracy of spectra generated using the
``cosmic confusion'' \cite{Bondetal} assumption to those calculated using
numerical evolution of the Boltzmann hierarchy.  For a range of parameters
of current interest the spectra are found to agree to $\sim10\%$ for all
multipoles up to the first peak ($\ell\simeq220$).
Over the same range the height of the peak is reproduced to $\simlt5\%$.
As pointed out by Dodelson \& Kosowsky \cite{DodKos}, the presence of
this degeneracy for scales greater than $0\degrees5$ allows a simple
statement of the parameter constraints arising from CMB data.  Given
current data we find that $\nu$ (see Eq.~\ref{eqn:nudef}) is constrained
to be $\nu=-0.05\pm0.07$.
By limiting the amount of spectral tilt, this result in combination with
large-scale structure measurements (on somewhat smaller scales) strongly
constrains models of structure formation (see e.g.~\cite{SSW,WSSD,WhiSco}
and \S\ref{sec:fit}).

\acknowledgements
I would like to thank Wayne Hu for conversations.



\begin{table}
\begin{tabular}{cc|cc|cc}
$\ell$ & $D_\ell$ & $\ell$ & $D_\ell$ & $\ell$ & $D_\ell$ \\ \hline
   2 &  0.91 &   54 &  1.63 &  178 &  4.63 \\
   4 &  0.92 &   64 &  1.78 &  195 &  4.93 \\
   6 &  0.95 &   74 &  1.94 &  212 &  5.07 \\
  10 &  1.00 &   84 &  2.12 &  230 &  5.03 \\
  14 &  1.06 &   96 &  2.38 &  248 &  4.80 \\
  19 &  1.13 &  108 &  2.67 &  267 &  4.39 \\
  25 &  1.22 &  120 &  3.00 &  287 &  3.87 \\
  31 &  1.31 &  134 &  3.41 &  308 &  3.29 \\
  38 &  1.41 &  148 &  3.83 &  329 &  2.79 \\
  46 &  1.52 &  163 &  4.26 &  351 &  2.42
\end{tabular}
\caption{The multipole moments, $D_\ell\equiv\ell(\ell+1)C_\ell$, for
the ``standard'' Cold Dark Matter model with $\Omega_0=1$, $\Omega_B=0.05$,
$n=1$ and $C_2^{(T)}=0$.  The moments have been normalized to have
$D_{10}=1$.  Enough moments have been given to interpolate the behaviour of
the curve.}
\label{tab:scdm}
\end{table}

\begin{table}
\begin{tabular}{c|c|c|c|c|c}
$\Omega_0$ & $h$ & $n$ & $\Omega_B$ & $z_R$ & $\tau$ \\ \hline
 0.3 &  0.90 & 0.92 & 0.02 & 28 & 0.165 \\
 0.4 &  0.70 & 0.93 & 0.03 & 23 & 0.141 \\
 0.5 &  0.57 & 0.94 & 0.05 & 20 & 0.126 \\
 0.6 &  0.49 & 0.95 & 0.06 & 18 & 0.115 \\
 1.0 &  0.45 & 0.80 & 0.10 & 11 & 0.065 \\
 1.0 &  0.50 & 1.00 & 0.05 & 45 & 0.271 \\
\end{tabular}
\caption{A ``reasonable'' estimate of the ionization redshift, $z_R$, for
some models.  The optical depth $\tau$ from $z=0$ to $z_R$ assuming full
ionization of the hydrogen is also shown.
The models with $\Omega_0<1$ have been chosen to be ``best fits'' to LSS data.
Specifically they have $\Omega_Bh^2=0.015$, a ``shape parameter''
$\Gamma=0.25$ and $\sigma_8=0.7\Omega_0^{-0.6}$ (which fixes $n$).
Since $n<1$ the effective shape measured by the galaxy correlation function
will be $\Gamma_{\rm eff}\simeq0.21$--$0.24$.
The last two models are a ``best-fitting'' and a ``standard'' CDM model.}
\label{tab:zr}
\end{table}


\end{document}